\documentclass[runningheads]{llncs}
\usepackage{graphicx}
\usepackage{mathtools}
\usepackage{listings}
\usepackage{proof}
\usepackage{amsfonts}
\usepackage{amssymb}
\usepackage{stmaryrd}
\usepackage{algorithm,algorithmic}
\usepackage{tabularx}

\newcommand{\floor}[1]{{\lfloor}#1{\rfloor}}
\newcommand{\braces}[1]{{\{}#1{\}}}

\begin{document}
\title{Leveraging Self-Sovereign Identity in Decentralized Data Aggregation\thanks{This is the preprint version of the conference paper "Leveraging Self-Sovereign Identity in Decentralized Data Aggregation" (Best Student Paper Award) in \textit{Proc. International Conference on Software Defined Systems (SDS), 2022}.}}
\titlerunning{Leveraging Self-Sovereign Identity in Decentralized Data Aggregation}
%
\author{Yepeng Ding\inst{1,2}\orcidID{0000-0002-6996-9333} \and
Hiroyuki Sato\inst{1} \and Maro G. Machizawa\inst{2}}
\authorrunning{Y. Ding et al.}
%
\institute{The University of Tokyo, Tokyo, Japan \\
\email{\{youhoutei,schuko\}@satolab.itc.u-tokyo.ac.jp} \and
Hiroshima University, Hiroshima Japan \\
\email{machizawa@hiroshima-u.ac.jp}
}
\maketitle              
\begin{abstract}
Data aggregation has been widely implemented as an infrastructure of data-driven systems. However, a centralized data aggregation model requires a set of strong trust assumptions to ensure security and privacy. In recent years, decentralized data aggregation has become realizable based on distributed ledger technology. Nevertheless, the lack of appropriate centralized mechanisms like identity management mechanisms carries risks such as impersonation and unauthorized access. In this paper, we propose a novel decentralized data aggregation framework by leveraging self-sovereign identity, an emerging identity model, to lift the trust assumptions in centralized models and eliminate identity-related risks. Our framework formulates the aggregation protocol regarding data persistence and acquisition aspects, considering security, efficiency, flexibility, and compatibility. Furthermore, we demonstrate the applicability of our framework via a use case study where we concretize and apply our framework in a decentralized neuroscience data aggregation scenario.

\end{abstract}
\section{Introduction}
\label{sec:intro}
Data aggregation is an essential process for compiling information from multiple data sources. It is commonly used in sensor networks \cite{krishnamachari_impact_2002,he_pda_2007,patel_data_2012,boubiche_big_2018} and Internet of Things (IoT) systems \cite{rahman_comparison_2016,pourghebleh_data_2017,yousefi_data_2021}, distributed data analysis \cite{salloum_random_2019} and machine learning \cite{xing_petuum_2015,mai_kungfu_2020,verbraeken_survey_2020}. In distributed environments, data sources are usually scattered randomly and constructed with heterogeneous architectures according to different specifications. Besides, any physical entity can provide data sources, such as individuals, organizations, programs, and devices. Consequently, data aggregation models are elaborated in distributed systems to ensure the controllability, interoperability, security, and privacy of gathering authorized data from authenticated data sources.

A typical data aggregation model has a centralized architecture where the control plane manages data streams provided by a set of data sources with a common interface and directs the orchestrated stream to a consumer, e.g., a device, server, or data center, which is constructed upon a set of trust assumptions as follows.
\begin{enumerate}
    \item[$a_1$] The control plane trusts that data sources ensure data security, including authenticity and availability.
    \item[$a_2$] Data sources trust that the control plane assigns correct privileges to consumers to access the requested data.
    \item[$a_3$] The control plane and data sources trust that communication protocols preserve data privacy.
    \item[$a_4$] Data sources and consumers retain anonymity to the control plane.
\end{enumerate}

However, designing and implementing a mechanism to satisfy these trust assumptions is challenging, especially while building a zero trust security model for data aggregation. For $a_1$, while an audit system can enhance the trust of internal data sources, i.e., the control plane and data sources are controlled by the same party, external data sources can still manipulate data before streaming to the control plane. $a_2$ depends on the access control provided by the control plane, which faces vulnerabilities and threats widely contained in centralized systems, such as single point of failure and denial of service attacks. Although $a_3$ can be theoretically satisfied by data encryption, the key exchange and management still face issues that can lead to data leakage. Moreover, $a_4$ is highly possible to contradict $a_1$ and $a_2$ because the trust establishment relies on bidirectional authentication and identity proving.

In recent years, distributed ledger technologies (DLTs) have provided a way to circumvent the trust assumptions above. Based on DLTs, a data aggregation model can function without centralized mechanisms, which we call a decentralized data aggregation model. In this model, the control plane is decentralized and controlled by a set of arbitrary nodes in a network. These nodes agree on an execution result of some control plane functionality based on a consensus mechanism such as proof of work, PBFT, and Raft. Data sources and consumers are separate from the centralized management of the control plane and participate in the data aggregation process via the reachable nodes. Besides, the rise of decentralized data persistence and sharing techniques \cite{ding_dagbase_2020,hu_ghostor_2020,ding_derepo_2020,hoang_privacy-preserving_2020,ding_sunspot_2021} are promising to resolve data security and privacy issues during data acquisition. However, the lack of identity management is vulnerable to impersonation attacks, data source manipulation, and collusion, potentially violating $a_1$ and $a_2$. Therefore, the main problem shifts from the assurance of security and privacy of data to the soundness of identity management.

With the advancement of self-sovereign identity (SSI) \cite{muhle_survey_2018}, an emerging identity model that enables physical entities to control their identity information, authentication and authorization verification that retain partial anonymity and total pseudonymity without centralized mechanisms become realizable, which forms a decentralized identity management mechanism \cite{ding_self-sovereign_2022}. Physical identities register their decentralized identifiers (DIDs) \footnote{https://www.w3.org/TR/did-core/} into a verifiable data registry (VDR) and associate the verifiable credentials (VCs) \footnote{https://www.w3.org/TR/vc-data-model/} issued and endorsed by authorities with their DIDs. In this manner, physical entities can prove their credentials to third parties by presenting the corresponding VCs, and third parties can verify the authenticity and endorsement of the presented credentials without relying on any middleware. Unfortunately, few intelligent frameworks integrate SSI into decentralized data aggregation to address potential authentication and authorization issues.

In this paper, we formulate a decentralized data aggregation framework to solve security and privacy issues related to the trust assumptions of centralized models through a fine-tuned SSI model. We summarize our main contributions as follows.

\begin{itemize}
    \item We present a formalized decentralized data aggregation framework integrating an SSI model to lift and eliminate the necessity of the trust assumptions $\braces{a_1, a_2, a_3, a_4}$.
    \item We elaborate aggregation protocols regarding two aspects: data persistence and data acquisition. We formulate a decentralized storage approach to implement a fully decentralized mechanism for data persistence and two types of data acquisition approach addressing security and efficiency, respectively.
    \item We concretize and apply our framework in a decentralized neuroscience data aggregation system to demonstrate its applicability.
\end{itemize}

\section{Related Work}
Decentralized data aggregation has a growing interest in recent years \cite{chen_when_2018,fan_consortium_2019,xie_decentralized_2021}. In work \cite{chen_when_2018}, \textit{LearningChain} is proposed to mechanize a decentralized data aggregation mechanism to serve the decentralized stochastic gradient descent algorithm. Although the authors demonstrate the effectiveness through experiments, this ad hoc framework is unsuitable for general data aggregation purposes. The authors of \cite{fan_consortium_2019} propose \textit{CBSG}, a decentralized data aggregation framework for the smart grid based on a consortium blockchain, of which the applicability is further improved by the work \cite{xie_decentralized_2021}. However, these works lack identity management mechanisms and require expensive on-chain interactions for aggregation processes.

Decentralized data sharing \cite{hu_ghostor_2020,ding_sunspot_2021} is a general form of decentralized data aggregation. The work \cite{hu_ghostor_2020} proposes \textit{Ghostor} to provide anonymity and verifiable linearizability based on decentralized trust. In work \cite{ding_sunspot_2021}, a decentralized framework called \textit{Sunspot} is proposed to enable privacy-preserving data sharing across public blockchains. However, extra efforts are required to concretize them as aggregation mechanisms with authentication and access control.

SSI is a rapidly developing research field and conceptually promising as an alternative to typical DLT-enabled authentication and access control frameworks like \textit{Bloccess} \cite{ding_bloccess_2023}. In recent years, there have been many studies on applying SSI to lift trust assumptions in centralized identity models \cite{bartolomeu_self-sovereign_2019,houtan_survey_2020}. To further improve the usability of SSI models, the work \cite{ding_self-sovereign_2022} formulates self-sovereign identity as a service. However, adapting SSI models to system contexts is indispensable. To the best of our knowledge, SSI has not been studied as the identity model for decentralized data aggregation.

\section{Decentralized Data Aggregation Framework}
\label{sec:framework}
\begin{figure}[htbp]
\centerline{\includegraphics[scale=0.39]{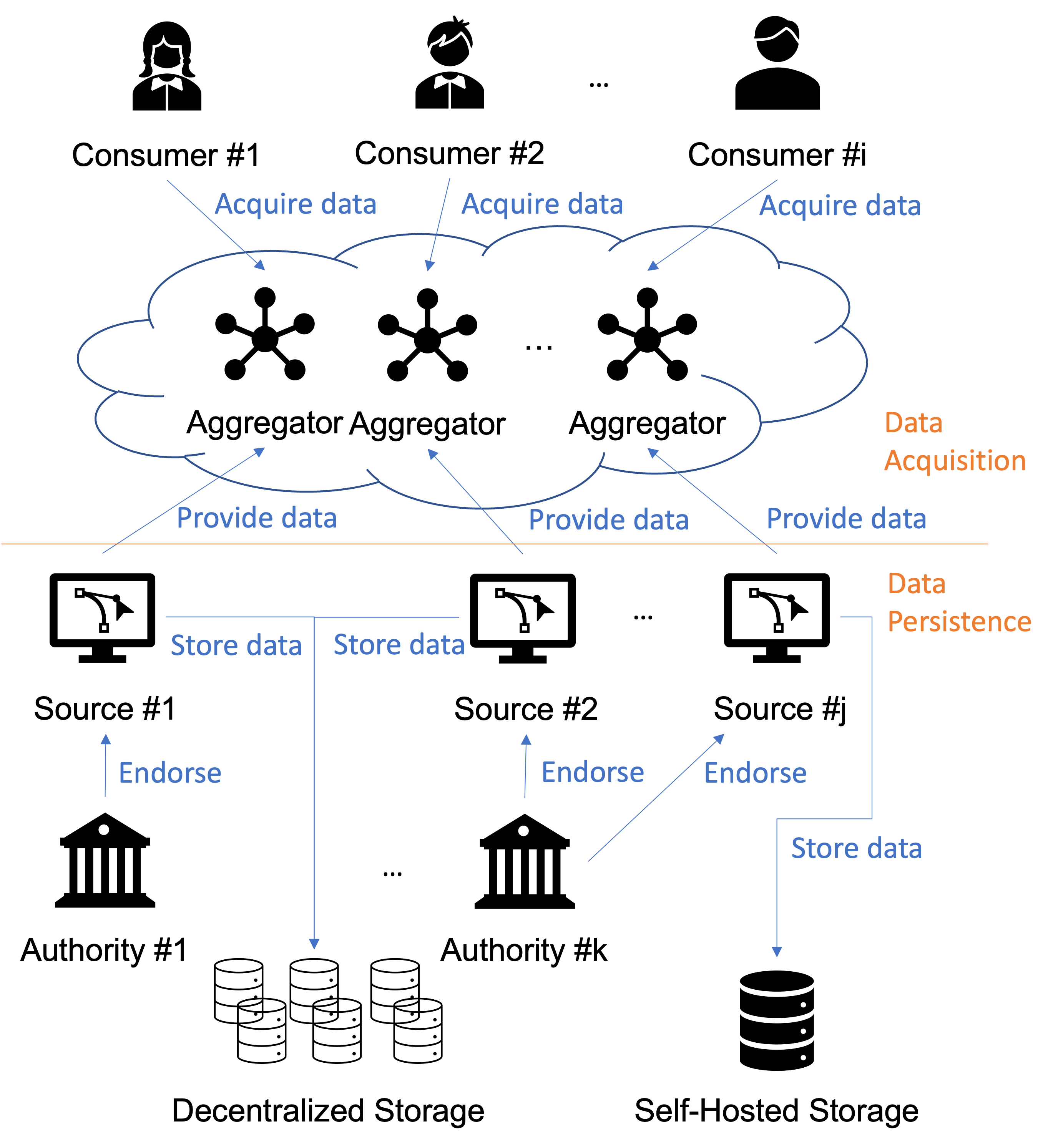}}
\caption{The conceptual diagram of our framework. The blue labels attached to the arrows represent the conceptual relation between the elements at two ends. The orange horizontal division line divides the diagram into two aspects labeled by orange texts.}
\label{fig:overview}
\end{figure}

Our framework considers two aspects of the data aggregation process under the SSI scheme: \textit{data persistence} and \textit{data acquisition}, of which the conceptual diagram is depicted in Fig~\ref{fig:overview}. 

An aggregator is a hybrid application partially deployed on a distributed ledger as a decentralized application, which bridges authentication, authorization, as well as data storage, collection, and processing. Consumers interact with aggregators to obtain on-demand data, while data sources provide data and meta information to aggregators. Notably, authorities, a new role we introduce to endorse and authorize the data provided by data sources, also interacts with aggregators during the endorsement process.

Although both \textit{data persistence} and \textit{data acquisition} aspects are centered around aggregators, \textit{data persistence} is the viewpoint of authorities and sources, and \textit{data acquisition} stands on the perspective of consumers.

The \textit{data persistence} aspect focuses on facilitating data management for the provider side. To fully realize decentralization in data aggregation, we elaborate a decentralized data storage approach that allows data sources to manage their data scattered on a set of nodes even not controlled by them in a decentralized network. Since our framework only recognizes one address, instead of a set of data storage locations, per data source, a data source can get the complete data by acquiring data partitions from a collection of storage nodes. Notably, this differs from \textit{data acquisition}, where consumers request to acquire data managed by multiple data sources. Besides, our framework has the flexibility to be compatible with self-hosted storage systems.

The \textit{data acquisition} aspect enables operationalizing the process of on-demand data collection from multiple data sources for consumers. An aggregator constructs connections with verified data sources, unifies data formats according to provided specifications, and combines the collected data.

\subsection{Formalization}
\begin{definition}[Distributed Ledger]
\label{def:dl}
A distributed ledger $\mathfrak{L}$ is composed of a finite node set $N$ where $|N| > 1$. Given a transaction $x \in X$, $x$ is accepted at node $n \in N$ if $\mathcal{F}(x, n) = \top$, where $\mathcal{F}: X \times N \mapsto \{ \top, \bot \}$ and $X$ is the submitted transaction set. Transaction $x$ is finalized in the network if $|\{ n \mid \mathcal{F}(x, n) = \top, n \in N \}| > \delta |N|$ where $\delta$ is the threshold for the network to reach consensus. The finalized transaction set is denoted by $\hat{X}$.

Additionally, we define the transaction property access operation as $[*]$, where $*$ can be any valid property name.
\end{definition}

\begin{example}[Threshold]
If $\mathfrak{L}$ adopts the PBFT as its consensus mechanism, then $\delta = \frac{2}{3}$. If the consensus mechanism is the proof of work, then $\delta = \frac{1}{2}$.  
\end{example}

\begin{definition}[Roles]
The set $O$, $C$, $S$ denote the authority set, consumer set, and data source set, respectively, where $O \neq \emptyset$ and $C \neq \emptyset$. The data source set $S$ satisfies that $|S| \geq |O|$. $\forall s \in S$, $s$ is endorsed by an authority $\mathcal{A}(s)$ where $\mathcal{A}: S \mapsto O$. An authority $o \in O$ endorses $\mathcal{A}^*(o)$ data sources where $\mathcal{A}^*: O \mapsto \wp(S)$.

Note that a physical entity can have multiple roles, each of which is a virtual entity of that physical entity. The set $E = O \cup C \cup S$ is the complete set of virtual entities of all roles. We shall omit \textit{virtual} in the rest of the paper.
\end{definition}

\begin{definition}[Key and Cryptography]
\label{def:keys}
Any entity and aggregator can generate a symmetric key $\kappa$.

$\forall e \in E$, e can generate a pair of asymmetric keys $(\textit{pk}_e, \textit{sk}_e)$ with identifier $\iota_e$ and derive a unique wallet address $\mathcal{W}(\textit{pk}_e)$ in $\mathfrak{L}$.

The signing function $\mathcal{S}_s$ takes an object $\textit{obj} \in \braces{0, 1}^*$ and a private key $\textit{sk}$ as the input, and outputs a signature $\mathcal{S}_s(\textit{obj}, \textit{sk})$.

The signature verification function $\mathcal{S}_v$ takes a signature $\mathcal{S}_s(\textit{obj}, \textit{sk})$ and a public key $\textit{pk}$ or (a key identifier) as the input, and outputs $\textit{obj}$.

The encryption function $\mathcal{E}_e$ takes an object $\textit{obj} \in \braces{0, 1}^*$ and a symmetric key $\kappa$ (or an asymmetric key $\iota$) as the input, and outputs a ciphertext $\mathcal{E}_e(\textit{obj}, \kappa)$ (or $\mathcal{E}_e(\textit{obj}, \iota)$).

The decryption function $\mathcal{E}_d$ takes a ciphertext $\mathcal{E}_e(\textit{obj}, \kappa)$ (or $\mathcal{E}_e(\textit{obj}, \iota)$) and a symmetric key $\kappa$ (or $\iota'$) as the input, and outputs $\textit{obj}$.

\end{definition}

\begin{definition}[Decentralized Identifiers]
\label{def:did}
$\forall e \in E$, $e$ has a unique DID that can be resolved to a DID document. For brevity, we use $e.a$ to denote the resolving function $\mathcal{R}(e)(a)$ where $\mathcal{R}: E \mapsto \mathbf{I} \cup \{ \varnothing \}, \mathcal{I} \in \mathbf{I}: \alpha_{did} \mapsto \eta_{did}$, $\alpha_{did}$ and $\eta_{did}$ are the property set and the value set of the DID document.

In the following discussions, we mainly consider $\alpha_{did} = \{ \textit{id}, \textit{auth}, \textit{assert} \}$, described in Table~\ref{tab:did_property}.

\begin{table}[htbp]
\caption{The description of $\alpha_{did}$.}
\begin{center}
\begin{tabularx}{\textwidth}{|l|X|}
\hline
\textbf{Property} & \textbf{Description}\\
\hline
\textit{id} & The DID of the DID subject in the context. \\
\hline
\textit{auth} & The key identifier referring to a pair of asymmetric keys to authenticate the DID subject in the context. \\
\hline
\textit{assert} & The key identifier referring to a pair of asymmetric keys to express claims by the DID subject in the context. \\
\hline
\end{tabularx}
\label{tab:did_property}
\end{center}
\end{table}

\end{definition}

\begin{definition}[Verifiable Credentials]
\label{def:vc}
We denote a VC as $\mathcal{V} \in \mathbf{V}$ where $\mathcal{V}: \alpha_{vc} \mapsto \eta_{vc}$, $\alpha_{vc}$ and $\eta_{vc}$ are the property set and the value set of the VC document.

$\forall e \in E$, $e$ has a set of VCs $\{ \mathcal{V} \mid \mathcal{V}(\textit{credentialSubject})[\textit{id}] = e.\textit{id} \}$ where the operation $[*]$ accesses the value of a sub-property.

In this paper, we mainly consider the scope of VC properties as $\alpha_{vc} = \braces{\textit{id}, \textit{issuer} \textit{credentialSubject}, \textit{proof}}$ where $\textit{credentialSubject} \triangleq \langle \textit{id}, \textit{claim} \rangle$ and $\textit{proof} \triangleq \langle \textit{key}, \textit{value} \rangle$. We describe $\alpha_{vc}$ in Table~\ref{tab:vc_property}.

\begin{table}[htbp]
\caption{The description of $\alpha_{vc}$.}
\begin{center}
\begin{tabularx}{\textwidth}{|l|X|}
\hline
\textbf{Property} & \textbf{Description}\\
\hline
\textit{id} & The identifier unambiguously referring to the VC in the context. \\
\hline
\textit{issuer} & The DID of the issuer issuing the VC in the context. \\
\hline
\textit{credentialSubject}[\textit{id}] & The DID of the subject associated with the claim of the VC in the context. \\
\hline
\textit{credentialSubject}[\textit{claim}] & The object containing a set of statements about the subject of the VC in the context. \\
\hline
\textit{proof}[\textit{key}] & The key used to sign the VC in the context. \\
\hline
\textit{proof}[\textit{value}] & The signature of the VC in the context. \\
\hline
\end{tabularx}
\label{tab:vc_property}
\end{center}
\end{table}

\end{definition}

\begin{definition}
[Verifiable Credential Proof]
\label{def:vc_proof}
Given an intact VC $\mathcal{V} \in \mathbf{V}$ issued by an authority $o \in O$, its property $\textit{proof}$ is assigned with $\textit{proof}[\textit{key}] = o.\textit{assert}$ and $\textit{proof}[\textit{value}] = \mathcal{S}_s(\mathcal{V} \ominus \textit{proof}, \textit{sk}^{\textit{assert}}_o)]$, where $\ominus: \mathbf{V} \times \alpha_{vc} \mapsto \bar{\mathbf{V}}$ is an operation to remove a property from a given VC.
\end{definition}

\subsection{Aggregator}
\label{sec:aggregator}
An aggregator consists of six components: \textit{Controller}, \textit{Connector}, \textit{Authenticator}, \textit{Processor}, \textit{Registry} and \textit{Mediator}, which is the core of our framework.

\begin{figure}[htbp]
\centerline{\includegraphics[scale=0.39]{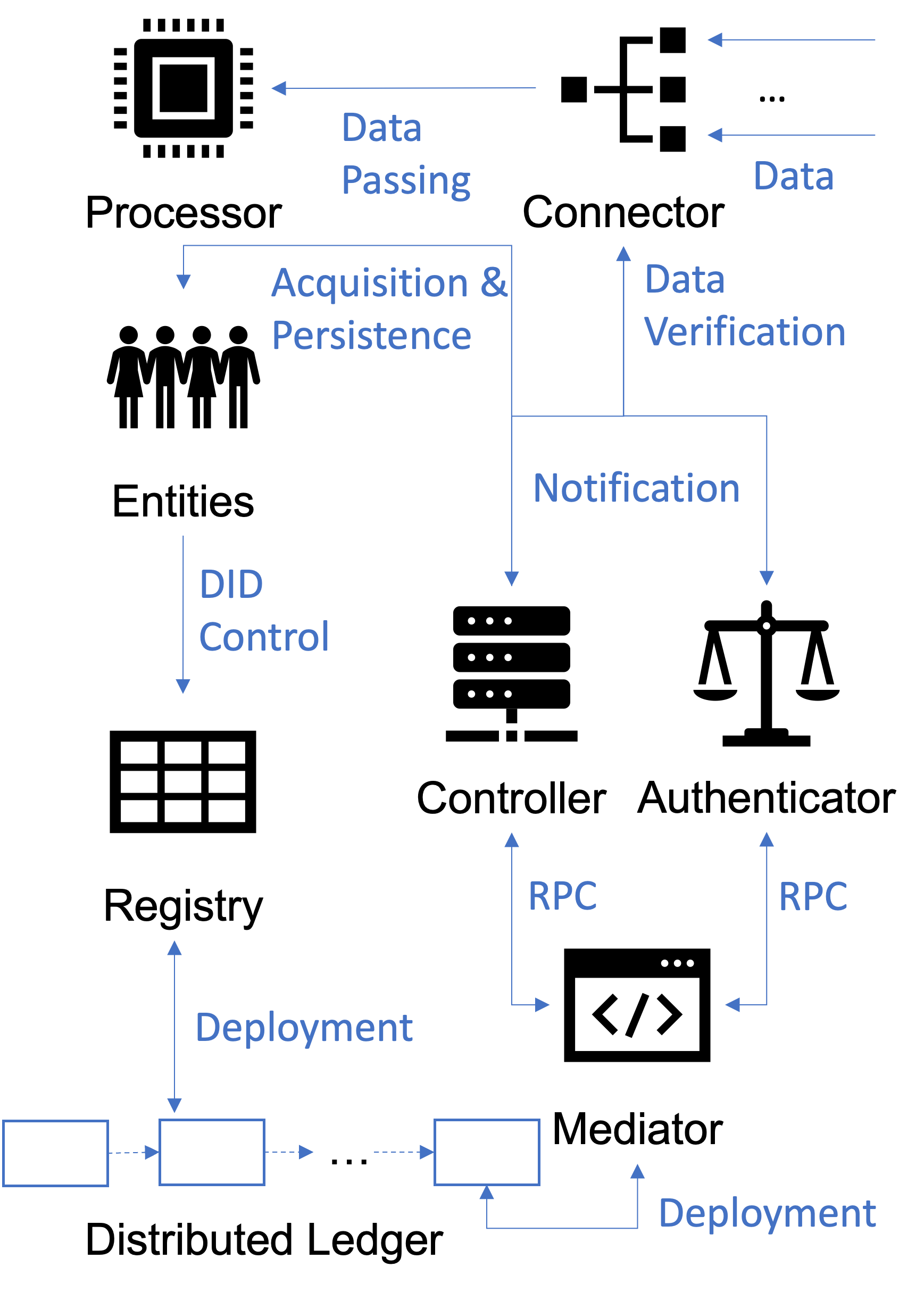}}
\caption{The architecture of the aggregator. The blue labels attached to the arrows represent the functional relation between the elements at two ends.}
\label{fig:aggregator}
\end{figure}

As shown in Fig~\ref{fig:aggregator}, entities interact with \textit{Controller} and \textit{Registry} for primary functionalities, including data acquisition, data persistence, and DID control. Although \textit{Controller}, \textit{Connector}, \textit{Authenticator} and \textit{Processor} are relatively independent components, an aggregator integrates them into a holistic system called \textit{Aggregator Client}. An \textit{Aggregator Client} does not rely on specific contexts, presenting the same functionalities to all entities. Consequently, each entity can have an \textit{Aggregator Client} in its local environment. Different from the \textit{Aggregator Client}, \textit{Registry} and \textit{Mediator} are decentralized components deployed on a distributed ledger, which theoretically has an infinite number of copies for entities. Therefore, countless copies of aggregators have the same behaviors in the network.

\subsubsection{Aggregator Client and Mediator}
An \textit{Aggregator Client} is a comprehensive system that handles the logic of data persistence and data acquisition with the support of \textit{Mediator} to interact with $\mathfrak{L}$.

Authorities can use \textit{Controller} to endorse the data provided by data sources. \textit{Controller} enforces endorsement propagation via RPCs on \textit{Mediator}. Besides, \textit{Controller} bridges communications between data sources and authorities.

Consumers interact with \textit{Controller} to make acquisition requests. \textit{Controller} communicates with data sources in both active and passive ways to exchange information (e.g., storage location) about the acquisition process and invoke RPCs (remote procedure calls) to read and write transactions in $\mathfrak{L}$ via \textit{Mediator}. \textit{Controller} also notifies \textit{Connector} when acquisition is ready. \textit{Connector} builds a set of connections with the given storage locations from \textit{Controller} to collect data in parallel with the assistant of \textit{Authenticator} to verify the collected data. \textit{Processor} is responsible for unifying the data format of the collected data according to a given specification.

We shall present the protocols of the \textit{data persistence} aspect in Section~\ref{sec:persistence}, and the \textit{data acquisition} aspect in Section~\ref{sec:acquisition}.

\subsubsection{Registry}
\label{sec:registry}
\textit{Registry} is a smart contract deployed on $\mathfrak{L}$ that manages the propagation, update, deletion, and resolve of DIDs for all roles.

\paragraph{DID Propagation}
According to Definition~\ref{def:did}, each entity has a DID propagating to be visible to the public.

\textit{Registry} provides the function \textit{Propagate} interface, usually an ABI (application binary interface), to enable the public visibility of its DID. The \textit{Propagate} function invocation appends a \textit{Propagation Transaction} $x: \tau_p \in \hat{X}$ to some block of $\mathfrak{L}$. A transaction of type $\tau_p$ contains the evaluation of $\alpha_{did}$ in the context of the invoking entity.

In this manner, DIDs are resolvable by searching $\tilde{X} = \{ x \mid x \in \hat{X}, x[\textit{did}] = e.id \}$. Notably, the search result $\tilde{X}$ is a set due to the possibility of DID updating and revoking. Therefore, the resolving function $\mathcal{R}$ defined in Definition~\ref{def:did} only considers the transaction $\tilde{x} \in \tilde{X}$ that $\forall x' \in \tilde{X} \setminus \tilde{x}: x'[\textit{timestamp}] < \tilde{x}[\textit{timestamp}]$ valid. If $\tilde{x}[\textit{deleted}] = \bot$, $\mathcal{R}$ parses $\tilde{x}$ to $\mathcal{I}_e$. Otherwise, $\mathcal{R}$ returns $\varnothing$.

\paragraph{DID Update}
Non-identifier properties such as $\textit{auth}$ and $\textit{assert}$ can be updated by invoking the \textit{Update} function of \textit{Registry}. The \textit{Update} function creates a new \textit{Update Transaction} $x: \tau_u \in \hat{X}$ in $\mathfrak{L}$. The $\tau_u$ transaction preserves the DID in $x[\textit{did}]$ and assigns the new evaluation of $\alpha_{did} \setminus \textit{id}$ to related transaction properties. $\varnothing$ is assigned to the deprecated property.

Therefore, $\mathcal{R}$ resolves a DID to its latest document by parsing the latest non-deleted transaction.

\paragraph{DID Deletion}
An entity $e$ can delete its propagated DID by invoking the \textit{Deletion} function of \textit{Registry}. A new \textit{Deletion Transaction} $x: \tau_d$ gets finalized after the invocation. A $\tau_d$ transaction $x$ satisfies $x[\textit{did}] = e.id \land x[\textit{deleted}] = \top$.

\subsection{Data Persistence}
\label{sec:persistence}
This section will present two types of data storage approaches for data sources after presenting persistence-related concepts. We elaborate a decentralized data storage approach in our framework to prevent potential central entities from weakening security, which is also the default approach. Our framework also provides flexibility for data sources to adopt the self-hosted storage approach. For both approaches, data sources are mandated to create semi-structured specifications and adapters for their data and include the specification in the data. For brevity, we use data to refer to data and its specification.

\subsubsection{Concepts}
\paragraph{Data Specification}
A data specification is semi-structured data that defines fields and types inside the data. For instance, a JSON-style data specification of personal data can be defined as \textit{\braces{"firstName": "string", "lastName": "string", "age": "number"}}. The nested structure is also allowed to create complicated data specification like \textit{\braces{"Name": \braces{"firstName": "string", "lastName": "string"}, "Person": \braces{"name": "Name", "age": "number"}}}.

When the raw data does not match the defined structure in a given data specification, the adapter reconstructs the raw data to ensure the provided data exactly matches the defined structure.

\paragraph{Data Endorsement}
A data source $s$ can ask $o \in O$ to endorse the provided data $d \in \braces{0, 1}^*$. If $o$ approves to endorse $d$, $o$ issues a VC through \textit{Controller} of an \textit{Aggregator Client} to $s$ as the ownership proof.

\begin{lemma}[Ownership]
\label{lemma:ownership}
A VC $\mathcal{V}$ provided by $s \in S$ implies $s$'s ownership of data $d$ if and only if $\mathcal{V}$ satisfies

\begin{enumerate}
    \item $\mathcal{V}(\textit{credentialSubject})[\textit{id}] = s.id$,
    \item $\mathcal{V}(\textit{credentialSubject})[\textit{claim}] = \mathcal{H}(d)$, and
    \item $\mathcal{V} \ominus \textit{proof} = \mathcal{S}_v(\mathcal{V}(proof)[\textit{value}], \mathcal{V}(proof)[\textit{key}])$,
\end{enumerate}
where $\mathcal{H}: \braces{0, 1}^* \mapsto \braces{0, 1}^l$ is a cryptographic hash function with a fixed length $l$.

\begin{proof}
According to Definition~\ref{def:vc}, $\mathcal{V}(\textit{credentialSubject})[\textit{id}]$ represents the DID of the subject associated with $\mathcal{V}(\textit{credentialSubject})[\textit{claim}]$. Hence, the first two equations $\mathcal{V}(\textit{credentialSubject})[\textit{id}] = s.id$ imply a credential statement that $s$ has $d$ identified by $\mathcal{H}(d)$. According to Definition~\ref{def:vc_proof}, if using the signature verification function $S_v$ in Definition~\ref{def:keys} $\mathcal{S}_v(\mathcal{V}(proof)[\textit{value}], \mathcal{V}(proof)[\textit{key}])$ can successfully recover the credential $\mathcal{V} \ominus \textit{proof}$, then the integrity of the credential is guaranteed by the asymmetric cryptography, which implies that the credential statement evaluates to be true, i.e., $s$ owns $d$.
\end{proof}

\end{lemma}

\subsubsection{Decentralized Storage}
Decentralized storage approach partitions and distributes data to a set of storage locations provided by incentivized third parties. Location sets are encapsulated into a transaction recorded into $\mathfrak{L}$. A data source can collect its stored data by reassembling data partitions collected from the parsed locations. Optionally, data sources can encrypt the data partitions before uploading them to storage locations to protect privacy.

A decentralized storage system has a layered architecture. It is composed of three layers: \textit{Partition Layer}, \textit{Mapping Layer}, and \textit{Chain Layer}.

\paragraph{Partition Layer}
\textit{Partition Layer} implements the data partition and assembly algorithm. Given data $d$ and scatter degree $\gamma \in [0, 1)$ as the input, the partition algorithm divides $d$ into a totally ordered set $D = \braces{d_0, \dots, d_i}$ where $i = \floor{\frac{1}{\gamma}}$ if $\gamma \neq 0$. Otherwise $d$ is partitioned into a set $\braces{d_0}$ where $d_0 = d$. The assembly algorithm is the reverse process of the partition algorithm. 

\paragraph{Mapping Layer}
\textit{Mapping Layer} maintains a dynamic table that records the available storage location set $\tilde{L}$. Given an ordered set $D$, \textit{Mapping Layer} assigns a location $l \in \tilde{L}$ to each $d \in D$ based on the business factors, such as availability, reputation, and storage cost. After creating the totally ordered location set $L$ for $D$, for all $d_i \in D$, $d_i$ is uploaded to the storage location $l_i \in L$. In the same way, \textit{Mapping Layer} fetches data partitions from a given $L$ and outputs the constructed data partition set.

\paragraph{Chain Layer}
\textit{Chain Layer} accepts a location set from \textit{Mapping Layer} and encapsulate it as a transaction $x: \tau_l \in \hat{X}$. Besides, given a transaction $x: \tau_l \in \hat{X}$, \textit{Chain Layer} parses $x$ into a location set.

Optionally, \textit{Chain Layer} uses cryptography to encrypt location sets to preserve the privacy of storage locations.

\subsubsection{Self-Hosted Storage}
\label{sec:self_hosted}
Data sources have high flexibility in selecting an appropriate location for data storage, because our framework does not directly interact with data storage locations. Therefore, a data source can host a database, data server, and software-defined data center as the storage location.

\subsection{Data Acquisition}
\label{sec:acquisition}
We formulate two data acquisition protocols: on-chain acquisition and off-chain acquisition. The on-chain acquisition elaborates DLT during protocol execution to enhance security, while the off-chain acquisition trades off certain security, especially availability, for higher efficiency and lower cost.

\subsubsection{Environment Setting}
We assume a consumer $c \in C$ with $c.\textit{id}$ propagated requests to aggregate the data from a propagated data source set $S' \subseteq S, |S'| > 1$ according to a transformation specification $\Psi$. $\forall s \in S'$, $s$ is endorsed by an authority $\mathcal{A}(s)$ via an ownership VC $\mathcal{V}_s$.

For simplicity, we only show the protocol executed by one data source. All other data sources follow the same protocol.

\subsubsection{On-Chain Acquisition}
\label{sec:on-chain}
We formalize the on-chain acquisition protocol as follows.
\begin{enumerate}
    \item $c$ makes an aggregation request containing $S'$ and $\Psi$ to the \textit{Controller} of $\mathfrak{A}_c$;
    \item The \textit{Controller} of $\mathfrak{A}_c$ invokes a RPC on the \textit{Mediator} to finalize a \textit{Collection Transaction} $x: \tau_c \in \hat{X}$ that $x[\textit{srcIds}] = \braces{s.id \mid s \in S'}$;
    \item $\forall s \in S$, $s$ requests $o = \mathcal{A}(s)$ to authorize $c$ if and only if $s.id \in x[\textit{srcIds}]$ for $x: \tau_c \in \hat{X}$;
    \item $o$ verifies the request by checking the existence of $x: \tau_c \in \hat{X}$ such that $s.id \in x[\textit{srcIds}]$;
    \item If $o$ rejects the request from $s$, then terminate the protocol. Otherwise, go into the next step;
    \item $o$ interacts with the \textit{Controller} of $\mathfrak{A}_o$ to finalize an \textit{Endorsement Transaction} $x: \tau_e \in \hat{X}$ such that $x[s] = s.id$ and $x[c] = c.id$;
    \item $s$ allocates a local space to collect data $d$ from the storage location, decrypt the data if applicable, re-encrypt the data with a symmetric key $\kappa$, and encrypt $\kappa$ by $\textit{pk}^{\textit{auth}}_c$;
    \item $s$ uploads $\mathcal{E}_e(d, \kappa)$ and $\mathcal{E}_e(\kappa, \textit{pk}^{\textit{auth}}_c)$ to a public storage space and obtains the storage information $m$;
    \item $s$ finalizes a \textit{Storage Transaction} $x: \tau_s \in \hat{X}$ such that $x[\textit{vc}] = \mathcal{E}_e(\mathcal{E}_e(\mathcal{V}_s, \textit{sk}^{\textit{auth}}_s), \textit{pk}^{\textit{auth}}_c)$ and $x[\textit{storage}] = \mathcal{E}_e(m, \textit{pk}^{\textit{auth}}_c)$ through the \textit{Controller} of $\mathfrak{A}_s$;
    \item The \textit{Connector} of $\mathfrak{A}_c$ parses $\braces{x \mid x: \tau_s \in \hat{X}}$ and dispatches $\mathcal{E}_e(\mathcal{V}_s, \textit{sk}^{\textit{auth}}_s)$ decrypted by $\textit{sk}^{auth}_c$ to the \textit{Arbitrator};
    \item The \textit{Arbitrator} of $\mathfrak{A}_c$ authenticates $s$ by verifying $\mathcal{E}_d(\mathcal{E}_e(\mathcal{V}_s, \textit{sk}^{\textit{auth}}_s), \textit{pk}^{\textit{auth}}_s)(\textit{credentialSubject})[\textit{id}] = s.id$, verifies the ownership according to Lemma~\ref{lemma:ownership}, and verifies the approval of $o$ by checking the existence of $x: \tau_e \in \hat{X}$ such that $x[s] = s.id \land x[c] = c.id$;
    \item If the \textit{Arbitrator} of $\mathfrak{A}_c$ fails to authenticate $s$ or verify the ownership, then terminate the protocol. Otherwise, go into the next step;
    \item The \textit{Connector} of $\mathfrak{A}_c$ fetches $\mathcal{E}_e(d, \kappa)$ and $\mathcal{E}_e(\kappa, \textit{pk}^{\textit{auth}}_c)$ based on $m$ decrypted by $\mathcal{E}_d(x[\textit{storage}], \textit{sk}^{\textit{auth}}_c)$;
    \item The \textit{Connector} of $\mathfrak{A}_c$ passes the data decrypted by $\mathcal{E}_d(\mathcal{E}_e(d, \kappa), \mathcal{E}_d(\mathcal{E}_e(\kappa, \textit{pk}^{\textit{auth}}_c), \textit{sk}^{\textit{auth}}_c))$ to the \textit{Processor};
    \item The \textit{Processor} of $\mathfrak{A}_c$ transforms $d$ based on $\Psi$ and returns the processed data to $c$.
\end{enumerate}

\subsubsection{Off-Chain Acquisition}
\label{sec:off_chain}
We formalize the off-chain acquisition protocol that improves efficiency and reduces the cost caused by interacting with $\mathfrak{L}$.
\begin{enumerate}
    \item $c$ makes an aggregation request containing $S'$, $\Psi$, and a random number $r \in \mathbb{N}$ to the \textit{Controller} of $\mathfrak{A}_c$;
    \item The \textit{Controller} notifies data sources in $S'$ with $r$ and public a port $z$ for the \textit{Connector};
    \item $\forall s \in S'$, $s$ requests $o = \mathcal{A}(s)$ to authorize $c$ with $\mathcal{E}_e(r, \textit{sk}^{\textit{auth}}_s)$;
    \item If $o$ rejects the request from $s$, then terminate the protocol. Otherwise, go into the next step;
    \item $o$ returns $\Omega = \mathcal{E}_e(\mathcal{E}_e(r, \textit{sk}^{\textit{auth}}_s), \textit{sk}^{\textit{auth}}_o)$ to $s$;
    \item $s$ allocates a local space to collect data $d$ from the storage location, decrypt the data if applicable, re-encrypt the data with a symmetric key $\kappa$, and encrypt $\kappa$ by $\textit{pk}^{\textit{auth}}_c$;
    \item $s$ uploads $\mathcal{E}_e(d, \kappa)$, $\mathcal{E}_e(\kappa, \textit{pk}^{\textit{auth}}_c)$ to a public storage space and obtains the storage information $m$;
    \item $s$ sends $\mathcal{E}_e(\mathcal{E}_e(\mathcal{V}_s, \textit{sk}^{\textit{auth}}_s), \textit{pk}^{\textit{auth}}_c)$, $\mathcal{E}_e(m, \textit{pk}^{\textit{auth}}_c)$, and $\Omega$ to the port $z$ of the \textit{Connector} of $\mathfrak{A}_c$;
    \item The \textit{Connector} of $\mathfrak{A}_c$ dispatches $\mathcal{E}_e(\mathcal{V}_s, \textit{sk}^{\textit{auth}}_s)$ decrypted by $\textit{sk}^{auth}_c$ and $\Omega$ to the \textit{Arbitrator};
    \item The \textit{Arbitrator} of $\mathfrak{A}_c$ authenticates $s$ by verifying $\mathcal{V}_s(\textit{credentialSubject})[\textit{id}] = s.id$, verifies the ownership according to Lemma~\ref{lemma:ownership}, and verifies the approval of $o$ by checking $\mathcal{E}_d(\mathcal{E}_d(\Omega, \textit{pk}^{auth}_{\mathcal{V}_s(\textit{issuer})}), \textit{pk}^{\textit{auth}}_s)$ where $\mathcal{V}_s = \mathcal{E}_d(\mathcal{E}_e(\mathcal{V}_s, \textit{sk}^{\textit{auth}}_s), \textit{pk}^{\textit{auth}}_s)$;
    \item If the \textit{Arbitrator} of $\mathfrak{A}_c$ fails to finish all types of verification, then terminate the protocol. Otherwise, go into the next step;
    \item The \textit{Connector} of $\mathfrak{A}_c$ fetches $\mathcal{E}_e(d, \kappa)$ and $\mathcal{E}_e(\kappa, \textit{pk}^{\textit{auth}}_c)$ based on $m$ decrypted by $\mathcal{E}_d(x[\textit{storage}], \textit{sk}^{\textit{auth}}_c)$;
    \item The \textit{Connector} of $\mathfrak{A}_c$ passes the data decrypted by $\mathcal{E}_d(\mathcal{E}_e(d, \kappa), \mathcal{E}_d(\mathcal{E}_e(\kappa, \textit{pk}^{\textit{auth}}_c), \textit{sk}^{\textit{auth}}_c))$ to the \textit{Processor};
    \item The \textit{Processor} of $\mathfrak{A}_c$ transforms $d$ based on $\Psi$ and returns the processed data to $c$.
\end{enumerate}

\section{Use Case}
\label{sec:use_case}
In this section, we show the concretization and application of our framework in a neuroscience data aggregation scenario to demonstrate its applicability.

Neuroscience data (e.g., electroencephalography, magnetic resonance imaging, and magnetoencephalography) collected from experiment participants are usually controlled by the institutes responsible for neuroscience experiments \cite{vaccarino_brain-code_2018}. Although participants have the privilege to access their data in most experiment agreements, sharing individual data with a third party is challenging in both technical and social aspects. Nevertheless, this has become an emerging demand for many reasons \cite{poldrack_making_2014}, such as maximizing the contribution of experiment participants, enhancing the reproducibility of neuroscientific research, providing a test bed for new methods, and reducing the cost of doing new analysis.

In our scenario, the requirement specification defines the main functionality of implementing a decentralized data aggregation system to facilitate neuroscience data sharing. The system makes experiment participants retain control of their data after experiments. In this manner, third parties can aggregate individual data by directly requesting participants. Participants can also actively share their data with third parties making data acquisition requests with the endorsement and permission of experiment institutes.

\subsection{Structure Refinement}
\begin{figure}[htbp]
\centerline{\includegraphics[scale=0.39]{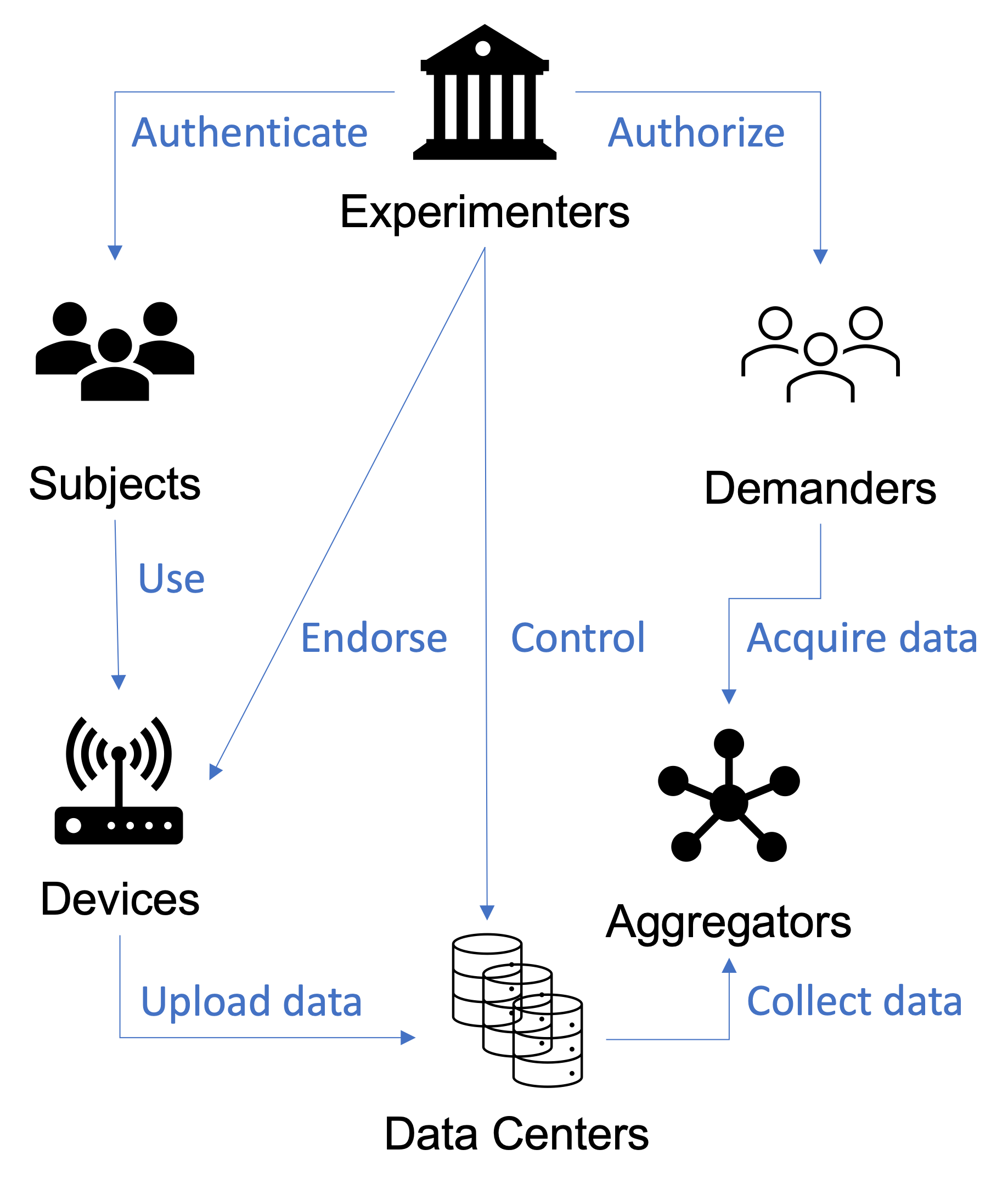}}
\caption{The conceptual diagram of the fine-tuned framework for decentralized neuroscience data aggregation. The blue labels attached to the arrows represent the conceptual relation between the elements at two ends. The interactions related to the \textit{Registry} of the \textit{Aggregator} are omitted.}
\label{fig:neuro_overview}
\end{figure}

To adapt our framework to the scenario, we extract three roles from the requirement specification: subject, experimenter, and demander. Subjects are experiment participants contributing data to neuroscientific research. Experimenters are entities collecting data from subjects, such as institutes and companies. Demanders are third parties making aggregation requests to acquire data. In general, these three roles can be mapped to the data source, authority, and consumer, respectively. We show the conceptual diagram in Fig~\ref{fig:neuro_overview}. Notably, we omit the interactions related to \textit{Registry}.

As shown in Fig~\ref{fig:neuro_overview}, authenticated subjects contribute their data through the devices endorsed by experimenters. Experimenters authorize demanders to acquire data by interacting with aggregators. Notably, we adopt data centers instead of the decentralized storage approach of our framework to preserve compatibility with most existing neuroscience data storage systems.

\subsection{Protocol Concretization}
We concretize the protocols shown in Section~\ref{sec:persistence} and Section~\ref{sec:acquisition} for this scenario through three phases: \textit{Initialization Phase}, \textit{Experiment Phase}, and \textit{Authorization Phase}.

\subsubsection{Initialization Phase}
Entities create their DIDs and propagate their DIDs by interacting with \textit{Registry} in the same way shown in Section~\ref{sec:registry} in \textit{Initialization Phase}.

\subsubsection{Experiment Phase}
\begin{figure}[htbp]
\centerline{\includegraphics[scale=0.39]{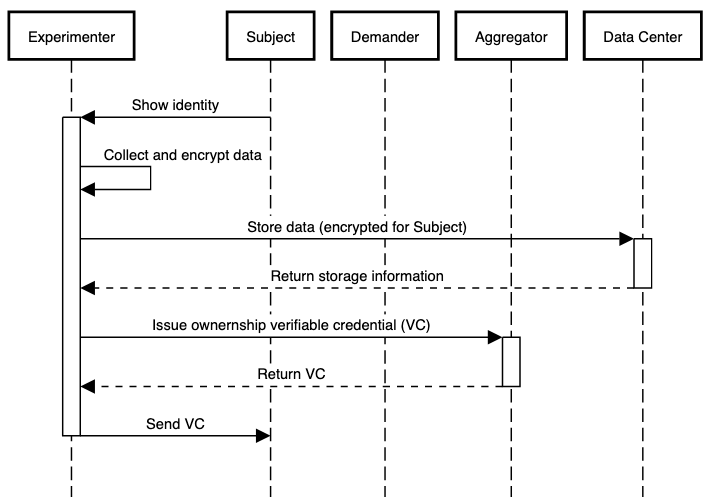}}
\caption{The sequence diagram of \textit{Experiment Phase} of the decentralized neuroscience data aggregation system.}
\label{fig:neuro_experiment}
\end{figure}

In this scenario, subjects use data centers provided by experimenters to store and manage the contributed data, which is the self-hosted approach introduced in Section~\ref{sec:self_hosted}. We show the sequence diagram of \textit{Experiment Phase} in Fig~\ref{fig:neuro_experiment}.

Notably, the encryption of the data collected from a subject $s$ is done by a symmetric key $\kappa$ that is later encrypted by $\textit{pk}_s$ with consideration of encryption performance in practice. The ownership VC follows the same specification in Section~\ref{sec:persistence}.

\subsubsection{Authorization Phase}
We formulate \textit{Authorization Phase} based on the off-chain aggregation illustrated in Section~\ref{sec:off_chain} regarding privacy protection and computation cost.

\begin{figure}[htbp]
\centerline{\includegraphics[scale=0.39]{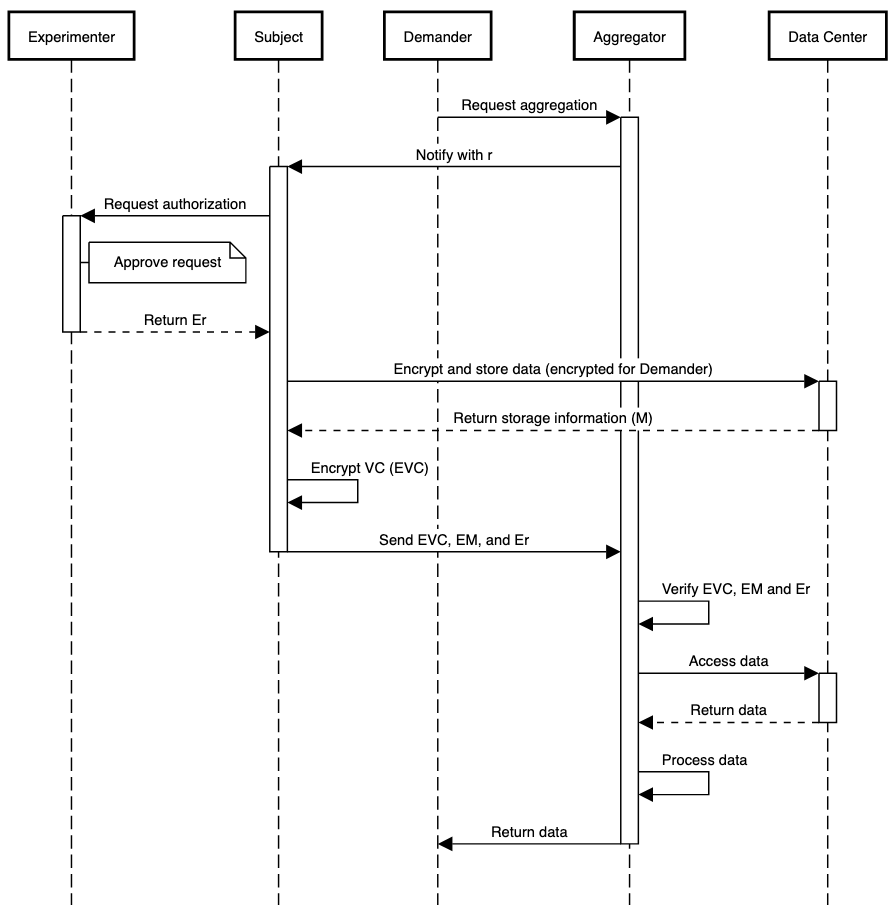}}
\caption{The sequence diagram of \textit{Authorization Phase} of the decentralized neuroscience data aggregation system. For simplicity, we only consider the case when an experimenter approves the authorization request.}
\label{fig:neuro_auth}
\end{figure}

As shown in the sequence diagram depicted in Fig~\ref{fig:neuro_auth}, a demander $c$ initializes an aggregation request sent to an aggregator. The aggregator notifies a subject $s$ with a random number $r \in \mathbb{N}$ to contribute the data by applying for authorization from the experimenter $o$ endorsing the data. If $o$ approves the request, $o$ encrypts $r$ in the same way as authorities and returns $\Omega$ in the protocol shown in Section~\ref{sec:off_chain}. $s$ re-encrypts the data by a new symmetric key $\kappa'$ and encrypts $\kappa'$ by $\textit{pk}_c$. The storage information $m$ is also encrypted by $\textit{pk}^{\textit{auth}}_c$. $s$ sends the encrypted ownership VC $\mathcal{E}_e(\mathcal{E}_e(\mathcal{V}_s, \textit{sk}^{\textit{auth}}_s), \textit{pk}^{\textit{auth}}_c)$, $\mathcal{E}_e(m, \textit{pk}^{\textit{auth}}_c)$, and $\Omega$ to the connection port. Then the aggregator follows the same protocol in Section~\ref{sec:off_chain} to verify the encrypted storage information and the ownership, collect and process data, and return the processed data to $c$.

\section{Discussion}
We argue that our framework has lifted or eliminated the trust assumptions of centralized data aggregation systems illustrated in Section~\ref{sec:intro}.

\begin{itemize}
\item[$a_1$]
To lift the trust assumption $a_1$, we introduce a new role called authority, of which the functionalities are refined from the concept \textit{issuer} in SSI. An authority acts as a trustworthy party for consumers by endorsing data sources. Based on Lemma~\ref{lemma:ownership}, malicious entities cannot deceive our framework by claiming illegitimate ownership and forging data. Ownership can always be proved by a VC sensitive to manipulation due to the carried proof.

Note that a data source can also endorse the data in our SSI scheme. In that case, $a_1$ still applies.

\item[$a_2$]
Our framework eliminates $a_2$ by leveraging the SSI mechanism that enables data sources to own and control data. As presented in Section~\ref{sec:acquisition} and the use case in Section~\ref{sec:use_case}, data sources and authorities entitle data access privileges for consumers. According to Definition~\ref{def:dl}, we can regard that access control integrity is ensured under the threshold assumption. 

\item[$a_3$]
For the protocols presented in Section~\ref{sec:framework}, we ensure all sensitive information is encrypted during communications to preserve data privacy. Particularly, the data to be aggregated remain encrypted during the whole aggregation process. Data can only be accessed by its sources before the aggregation. Data is also re-encrypted before being passed to the \textit{Connector} of aggregators.

In our use case study, experimenters may have access to the data contributed by subjects, which is a typical case for neuroscience experiments.

Hence, the assumption of $a_3$ is weakened as the assumption of cryptography correctness.

\item[$a_4$]
Data sources and consumers are scattered in a decentralized network, only whose DIDs are propagated to the framework, as illustrated in Section~\ref{sec:aggregator}. Data source identities may be known to authorities for endorsements, and consumer identities may be disclosed to data sources for authorization. Nevertheless, there is no direct way for uninvolved parties to disclose physical identities, i.e., $a_4$ is lifted by pseudonymity.
\end{itemize}

Furthermore, our decentralized architecture provides high availability, especially in the on-chain acquisition protocol. Although it is possible to disturb the normal functioning of data sources by exploiting the exposed physical network addresses, attackers must make considerable efforts to reveal the physical identities of data sources due to the lifted trust assumption $a_4$. Besides, a physical entity can easily relink an owned data source to a new physical network address to prevent malicious hunting.

\section{Conclusion}
In this paper, we have presented a decentralized data aggregation framework by leveraging SSI techniques. Our framework contains a set of data persistence and acquisition protocols to serve all types of roles involved in data aggregation. We have also presented the application of our framework in a decentralized neuroscience data aggregation system. Furthermore, we have discussed how our framework lifts and eliminates the trust assumptions in centralized data aggregation frameworks.

%
%
%
\bibliographystyle{splncs04}
\bibliography{references}
\end{document}